\documentclass[preprint,a4paper,superscriptaddress,aps,prl,onecolumn,floatfix,citeautoscript,showpacs]{revtex4}
\usepackage{graphicx}
\usepackage{epsfig}
\usepackage{latexsym}
\usepackage{amsmath,bm}

\usepackage{mciteplus}
\usepackage{natbib}
\newcommand{\kv}{\ensuremath{\bm{k}}}
\newcommand{\etal}{\textit{et al.~}}
\newcommand{\abinitio}{\textit{ab initio~}}
\newcommand{\gdir}{\ensuremath{E_{\textrm{g}}^{\textrm{dir}}}}
\newcommand{\gind}{\ensuremath{E_{\textrm{g}}^{\textrm{ind}}}}
\newcommand{\GW}{\ensuremath{GW}}
\newcommand{\GWz}{\ensuremath{G_0W_0}}

\begin{document}

\title{Effects of electronic and lattice polarization on the band structure of delafossite transparent conductive oxides}

\author{Julien Vidal}
\affiliation{Institute for Research and Development of Photovoltaic Energy (IRDEP), UMR 7174 CNRS/EDF/ENSCP, 6 quai Watier, 78401 Chatou, France}
\affiliation{Laboratoire des Solides Irradi\'es, \'Ecole Polytechnique, CNRS, CEA-DSM, 91128 Palaiseau, France}
\affiliation{European Theoretical Spectroscopy Facility (ETSF)}

\author{Fabio Trani}
\affiliation{LPMCN, Universit\'e Claude Bernard Lyon I and CNRS, 69622 Villeurbanne, France}
\affiliation{European Theoretical Spectroscopy Facility (ETSF)}

\author{Fabien Bruneval}
\affiliation{CEA, DEN, Service de Recherches de M\'etallurgie Physique, F-91191 Gif-sur-Yvette, France}
\affiliation{European Theoretical Spectroscopy Facility (ETSF)}

\author{Miguel A. L. Marques}
\affiliation{LPMCN, Universit\'e Claude Bernard Lyon I and CNRS, 69622 Villeurbanne, France}
\affiliation{European Theoretical Spectroscopy Facility (ETSF)}

\author{Silvana Botti}
\affiliation{Laboratoire des Solides Irradi\'es, \'Ecole Polytechnique, CNRS, CEA-DSM, 91128 Palaiseau, France}
\affiliation{LPMCN, Universit\'e Claude Bernard Lyon I and CNRS, 69622 Villeurbanne, France}
\affiliation{European Theoretical Spectroscopy Facility (ETSF)}

\begin{abstract}
We use hybrid functionals and restricted self-consistent \ensuremath{GW}, state-of-the-art theoretical approaches for quasiparticle band structures, 
to study the electronic states of delafossite Cu(Al,In)O$_2$, the first p-type and bipolar transparent
conductive oxides.
We show that self-consistent \ensuremath{GW} gives remarkably wider band gaps than all the other approaches used so far.
Accounting for polaronic effects in the \ensuremath{GW} scheme we recover a very nice agreement with experiments.
Furthermore, the modifications with respect to the Kohn-Sham bands are strongly \ensuremath{\bm{k}}-dependent,
which makes questionable the common practice of using a scissor operator.
Finally, our results support the 
view that the low energy structures found in optical experiments, and initially attributed to an indirect transition,
are due to intrinsic defects in the samples.
\end{abstract}


\maketitle

Many high-technology devices, such as flat panel displays, touch
screens, or even thin-film solar cells, require the use of thin
transparent contacts. These contacts are usually built from insulating
oxides that, for a certain range of doping, become conductive while
retaining transparency in the visible spectrum.  The most common
examples of these so-called transparent conductive oxides (TCOs) are
electron (n-)doped SnO$_2$, In$_2$O$_3$, and ZnO. Hole (p-)doping of
wide gap semiconductors was for long time very
hard to obtain~\cite{nakamura92,*neugebauer95}. It is therefore not surprising
that the discovery of p-doping in
CuAlO$_2$ thin films with a carrier mobility of about 10\,cm$^2$/(V\,s)
attracted great interest~\cite{benko84,*kawazoe97}.  Other members of the delafossite family,
like CuGaO$_2$~\cite{ueda01} and CuInO$_2$~\cite{yanagi01}, were
discovered shortly after. The latter compound is particularly
interesting as it exhibits bipolar (n- \textit{and} p-type) conductivity
by doping with appropriate impurities and tuning the
film-deposition conditions~\cite{yanagi01}. This opens the way to the
development of transparent p-n junctions, and therefore fully
transparent optoelectronic devices, functional windows and stacked solar cells
with improved efficiency.

CuAlO$_2$ is by far the most studied system of the family of
delafossite TCOs.
However, there is still no agreement either on the origin of the p-type conductivity,
or on the electronic bands of the pure crystal.
Measurements of the direct \textit{optical} band
gap (\gdir) of CuAlO$_2$ fall in the range from 2.9 to
3.9\,eV~\cite{benko84,*kawazoe97,yanagi00,ong03,*dittrich04,*alkoy05,
  *banerjee04,*banerjee05,*banerjee05-physicaB,*yu07,gilliland07,pellicer06},
with most values in the interval 3.4--3.7\,eV.  These
experiments also yield a large dispersion of indirect gaps (\gind),
from 1.65 to 2.1\,eV, with one experiment measuring
2.99\,eV~\cite{pellicer06}. Unfortunately, there is only one
photoemission experiment~\cite{yanagi00} that yields 3.5\,eV
for the \textit{quasiparticle} band gap. 
Note that the \textit{optical} and \textit{quasiparticle} gaps
differ by the exciton binding energy.
Concerning CuInO$_2$, optical
experiments measured \gdir\ between 3.9 and 4.45\,eV~\cite{yanagi01,teplin04,sasaki03}, with 
only one estimation of \gind\ at 1.44\,eV~\cite{sasaki03}. 

From the theoretical perspective, the situation is also quite complex,
even if the full Cu 3$d$ shell should exclude the strongly correlated electron regime.
These materials are usually studied within density functional theory (DFT), using the standard
local density (LDA) or generalized gradient approximations (GGA). 
However, it is well known that the Kohn-Sham band structures systematically underestimate the band gaps.
For similar compounds, like Cu$_2$O and CuIn(S,Se)$_2$, Kohn-Sham LDA
calculations lead to unreasonable band structures, in particular due
to the misrepresentation of the hybridization between the $d$
electrons of the metal and $p$ electrons of the
anion~\cite{bruneval06,vidal09}. To overcome this situation, hybrid
functionals have been recently proposed, with very
promising results~\cite{brothers08}, especially for materials with small and intermediate
band gaps~\cite{paier08,hafner08}. Other approaches include LDA+$U$, that tries to improve
the description of correlation through the introduction of a mean-field Hubbard-like term.
This method has been quite successful in the study of
strongly correlated systems, but it relies on a parameter $U$, that is often adjusted to experiments.

Arguably the most reliable and used \abinitio\ technique to obtain quasi-particle
band structures is the many-body \GW\ approach~\cite{hedin65}.
The common practice within this framework is to
start from a DFT calculation, and evaluate perturbatively the
\GW\ energy corrections to the band structure. This
procedure, which we will refer to as \GWz, is justified when
the departure wave functions and band structure are already close to
the quasiparticle ones. This is indeed the case in many systems, explaining 
why \GWz\ has been extremely successful in
describing electron addition and removal energies for metals,
semiconductors and insulators~\cite{aulbur00}.  However, it has been
recently shown that \GWz\ fails for many transition metal
oxides~\cite{bruneval06,gatti07}.

To solve this problem one can perform restricted self-consistent (sc)
\GW ~\cite{faleev04,*schilf06}. This technique has the
advantage of being independent of the starting point at the price of
large computational complexity. Fortunately, there is an alternative
procedure that yields wavefunctions that are extremely close to those
obtained in a full sc-\GW\ calculation, namely sc-COHSEX
as explained in Ref.~\cite{bruneval06-b}. The dynamical effects that are absent in  
COHSEX calculations can then be accounted for by performing a final 
perturbative \GW\ step. This method, that we will refer to as sc-\GW, 
has been applied to many oxide
compounds, yielding excellent results for the band gaps and the
quasiparticle band structure~\cite{bruneval06,bruneval06-b,gatti07,vidal09}.

\begin{figure}[t]
 \centering
 \includegraphics[width=0.9\columnwidth]{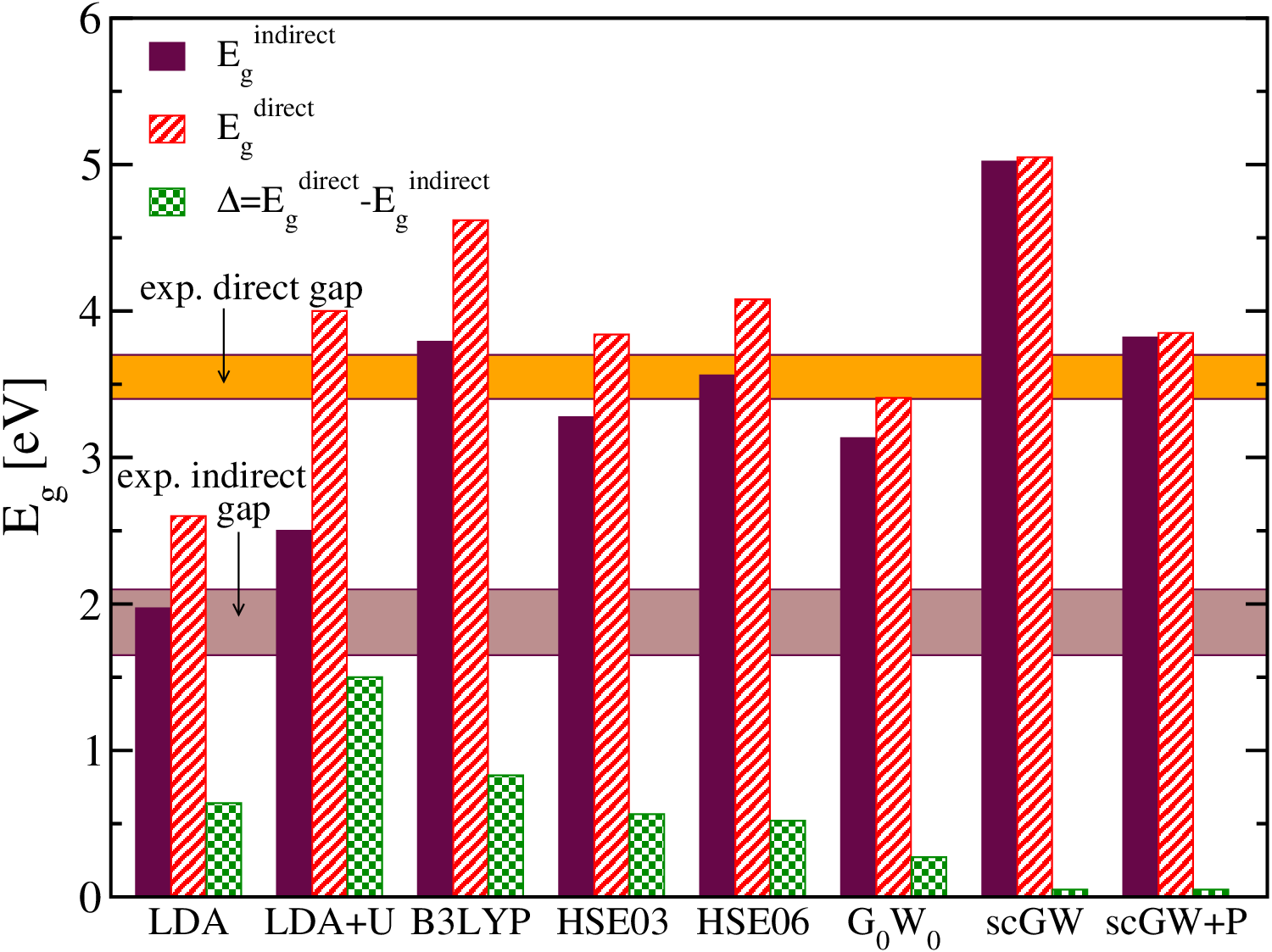}
 \caption{(Color online) Band gaps of CuAlO$_2$ using: LDA, LDA+$U$, hybrid, \GWz, sc-\GW, and sc-\GW\ including model
  polaronic corrections.
  The horizontal zones contain data extracted from various optical experiments (see text).}
\label{fig:fig1}
\end{figure}

Note that these theoretical techniques yield quasiparticle bands, and not optical gaps.
To evaluate these latter quantities one mostly resorts to the solution of the Bethe-Salpeter
equation. For the delafossite structures there is one such calculation starting from
a GGA+$U$ band structure~\cite{laskowski09}. It yields for CuAlO$_2$ a very large exciton binding energy of about 0.5\,eV 
for the first direct transition.
The choice of $U$ was found to have strong consequences on the width of the band gap, 
but it did not affect significantly the exciton binding energy. We can thus assume that 0.5\,eV is a reasonable
estimate.

\begin{figure}
 \centering		
\includegraphics[width=0.9\columnwidth]{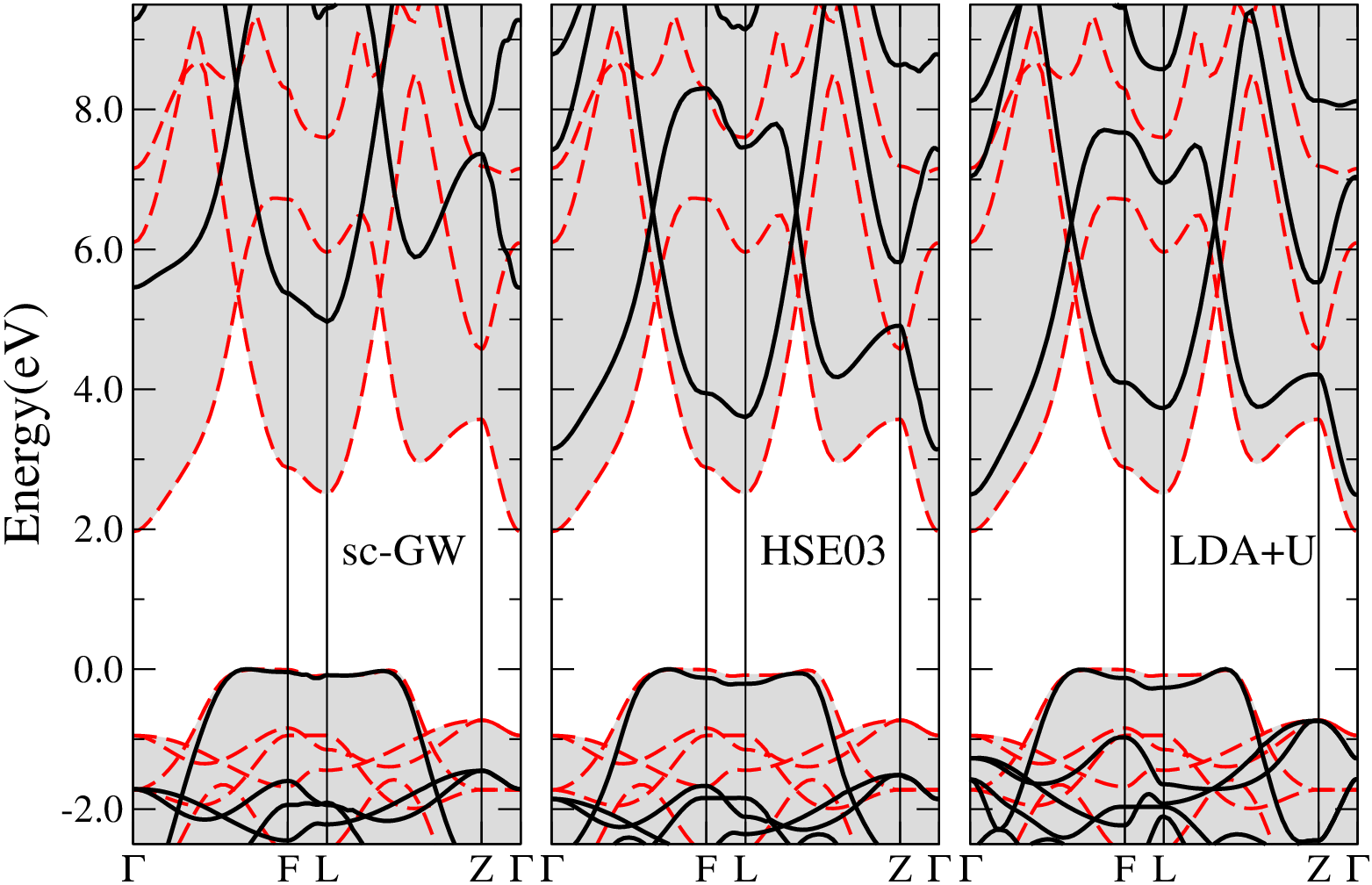}
\caption{(Color online) Band structures for CuAlO$_2$: comparison of LDA (red dashed lines) with sc-\GW\ (left panel),
HSE03 (central panel), and LDA+$U$ (right panel).}
\label{fig:fig2}
\end{figure}

In the following, we present calculations of the band structures of
CuAlO$_2$ and CuInO$_2$ using some of the most accurate theoretical
tools available in the community. These include the standard LDA, hybrid functionals
(namely B3LYP~\cite{B3LYP} and two flavors of Heyd-Scuseria-Ernzerhof, HSE03 and HSE06~\cite{HSE03,*HSE06}), LDA+$U$, \GWz\ and
sc-\GW.  As discussed above, we expect
sc-\GW\, to be the most accurate \abinitio\ approach. When the comparison was
possible, we found our results in excellent agreement with previous calculations (Refs.~\cite{yanagi00,
gilliland07,pellicer06,ingram01,*shi08,nie02,robertson02} for LDA, Ref.~\cite{robertson02} for B3LYP,
and Ref.~\cite{huda09} for GGA+$U$).

The hybrid and LDA+$U$ calculations were performed with VASP~\cite{vasp1,*vasp2} and ABINIT~\cite{abinit}
respectively, using the PAW formalism and an energy cutoff of 44\,Ha.
The parameter $U$ was set to 8\,eV as in Ref.~\cite{laskowski09}.
Our \GW\ calculations were performed with ABINIT, starting from
LDA band structures and using norm-conserving pseudopotentials with 
semicore states (3s and 3p for Cu and 4s and 4p for In) included in the valence. 
The energy cutoff was 120\,Ha for the ground state calculation, and 
the ${\bf k}$-point grid was a $4 \times 4
\times 4$ Monkhorst-Pack.
As the experimental and LDA relaxed geometries are very close (within 1\%),
and the small contraction of the lattice in LDA has a negligible effect
on band structures ($\le 0.05$\,eV), we employed experimental lattice parameters~\cite{robertson02}.
Note that it was absolutely essential to use the
method of Ref.~\cite{bruneval08}, due to the extremely slow
convergence with respect to the number of empty states.

In Figs.~\ref{fig:fig1} and ~\ref{fig:fig2} we show direct and indirect
photoemission gaps and the band structures of CuAlO$_2$ obtained using different
theoretical approaches.  The minimum \gdir\ of CuAlO$_2$ is always found at
L, where the dipole transition between the band edge states is
allowed~\cite{nie02}. All calculations, except sc-\GW, give a fundamental
\gind\ between the conduction band minimum at $\Gamma$ and the valence band maximum along
the $\Gamma$-F line.
The experimental data for optical gaps are also presented
with an error bar that reflects the dispersion of the most likely values found in
literature.
LDA exhibits, as expected, the smallest gaps. Basically every approach
beyond it opens up the gap by different amounts and modifies
the band dispersions. The direct and indirect gaps have
similar behaviors in the different theories, and both increase when
going from LDA$<$\GWz$<$HSE03$<$HSE06$<$B3LYP$<$sc-\GW. 
On the other hand, the difference $\gdir - \gind$ seems
to decrease with the sophistication of the method, reaching nearly
zero for the sc-\GW calculation.
This is a consequence of the drastic change 
of the conduction band dispersion, which displaces the conduction minimum from $\Gamma$ to L
when sc-\GW\ is applied (see Fig.~\ref{fig:fig2}). Only LDA+$U$ does not follow the trend, as it is the only case in which 
$\gdir - \gind$ gets significantly larger than in LDA.

\begin{figure}
 \centering
\includegraphics[width=0.9\columnwidth]{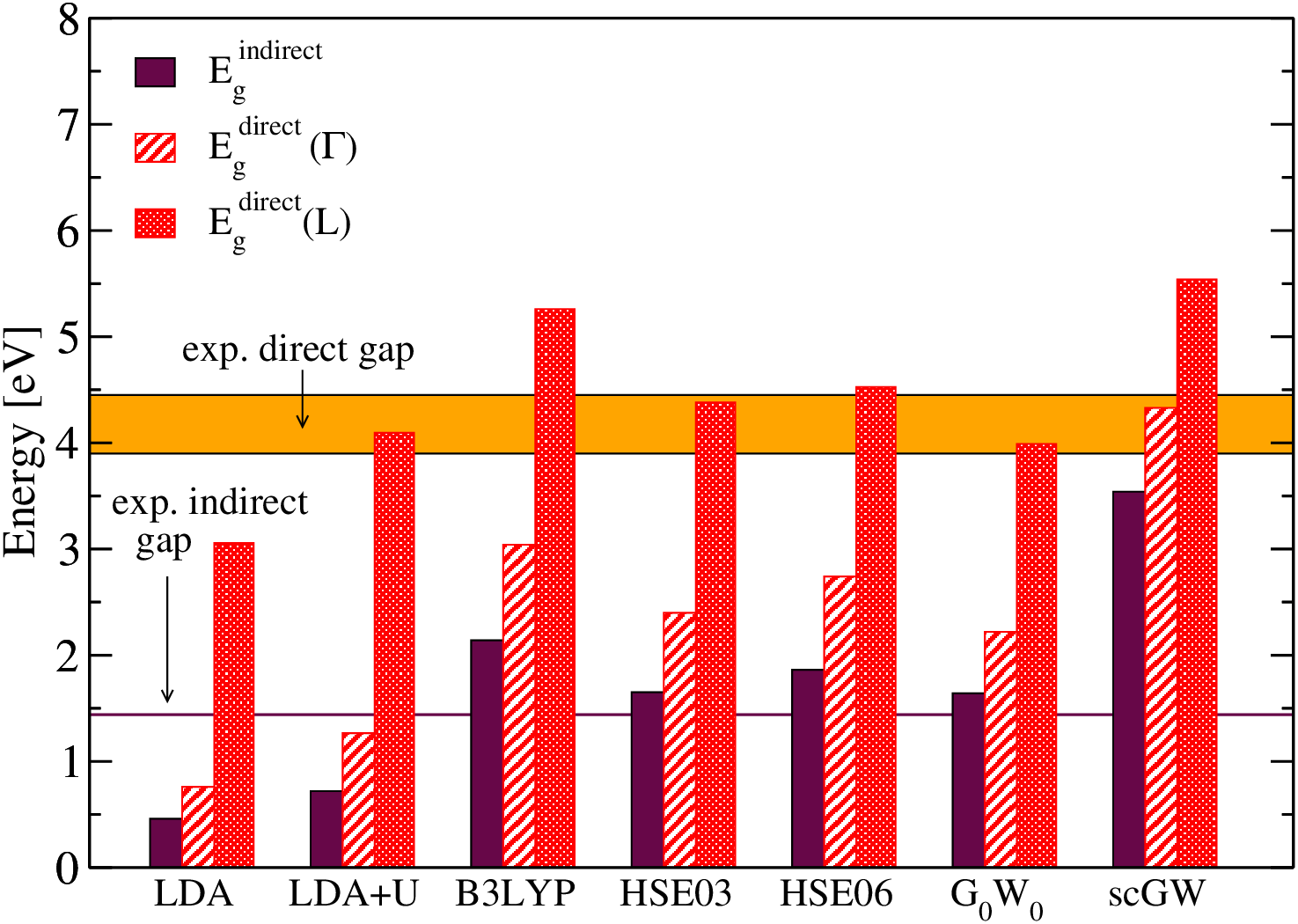}
\caption{(Color online) Band gaps of CuInO$_2$ using: LDA, LDA+$U$, hybrid, \GWz, sc-\GW.
  The horizontal zones contain data extracted from optical experiments (see text).}
\label{fig:fig3}
\end{figure}

Looking at the direct gap, we point out that most of the methods give results that are within the 
experimental range,  when an exciton binding energy of around 
0.5\,eV~\cite{laskowski09} is considered. This is true for LDA+$U$, \GWz, the hybrids HSE03 and HSE06.
However, for sc-\GW\ and even for B3LYP, the theoretical gap is larger
by about 1--1.5\,eV than the experimental findings.
For CuInO$_2$ (see Fig.~\ref{fig:fig3}) we
have to make the comparison with care, as the smallest \gdir\ is located at $\Gamma$,
where optical transitions are forbidden~\cite{nie02}.
A meaningful comparison with experiments must consider the gap at L. Thus, we find that both trends
and quantitative results are analogous to those for CuAlO$_2$. In particular, sc-\GW\ yields 
again \gdir\ larger by 1--1.5\,eV than the experimental range.

We stress again that, to date, sc-\GW\ is arguably the best method
available to estimate band gaps of wide-gap semiconductors, and that
it gives excellent results for compounds like Cu$_2$O and
CuIn(S,Se)$_2$~\cite{bruneval06,vidal09}.  It is unlikely that the
presence of defects can lead to such a large shrinkage of \gdir.
However, there is another effect that has been neglected up to now:
the change of screening due to the polarization of the lattice~\cite{Fowler1966,*Kunz1972}. In
fact, according to the experimental data~\cite{pellicer09},
unfortunately available only for CuAlO$_2$, the polaron constant
for this system is large ($\alpha_\text{p} \sim 1$), indicating a
non-negligible contribution of the lattice polarization to the
electronic screening. In other ionic compounds with
similar polaron constants this can lead to a shrinkage to the band gap
by about 1\,eV~\cite{bechstedt05}. A full sc-\GW\ calculation
including in an \abinitio\ framework the effects of the lattice
polarization is to date beyond reach.  However, a reliable estimate
can be obtained using the model proposed by Bechstedt
\etal \cite{bechstedt05}, which gives a static representation of the polaronic
effects based on difference of experimental static dielectric constants.
By performing a perturbative \GW\ step including model
polaronic effects on top of the sc-COHSEX, we found a
uniform (${\bf k}$-independent) shrinkage of the band gap by
1.2\,eV. As we can see in Fig.~\ref{fig:fig1}, this correction 
brings our results for \gdir\ well within the
experimental range (once the excitonic correction of about
0.5 eV is also considered). As it is observed in Ref.~\cite{bechstedt05},
the polaronic model employed can only overestimate the
correction. All these results point to the conclusion that
the agreement of the other methods with experiment was
fortuitous and due to a cancellation of errors.

Looking now at the indirect gap, we focus on Fig.~\ref{fig:fig1}
as there are more experimental data for CuAlO$_2$. All
the hybrids and \GW\ calculations yield indirect
gaps much larger than the experimental range 1.65–-2.1\,eV, 
even taking into account any possible excitonic
and polaronic effects. Moreover, sc-\GW, the best method
used in this work, yields the highest \gind\, at
around 5\,eV, while the difference $\gdir - \gind$ is in general
much smaller than the experimental value ($\approx$2\,eV), and
even vanishing for the sc-\GW\ calculation. From Fig.~\ref{fig:fig3} we
realize that these conclusions are as well valid for CuInO$_2$,
where the best estimates for the indirect band gap is
much larger than the experimental value of 1.44\,eV~\cite{sasaki03}.

These are very strong arguments in favor of Robertson
\etal~\cite{robertson02} that suggested that the experimental ``indirect
gap'' absorption was due to defects, and should not
be present in the defect-free compound. Also Pellicer-
Porres \etal~\cite{pellicer06} questioned the interpretation of the low
energy peaks as indirect transitions, as the absorption coefficient
is more than two orders of magnitude larger than
in typical indirect absorption edges. The most promising
defects are oxygen interstitials O$_\textrm{i}$, as
LDA calculations predict low formation
energies and the introduction of states in the gap at 0.7
and 1.4 eV~\cite{hamada06}. However, a full clarification of this issue
will require sc-\GW\ or hybrid calculations for these, and
other more complex defects.

Finally, we analyze more in detail the band structures
of CuAlO$_2$ shown in Fig.~\ref{fig:fig2}. LDA calculations (red dashed lines)
are compared with sc-\GW, HSE03, and
LDA+$U$ calculations. The main effect of LDA+$U$ is
to open the LDA gap by an amount that can be controlled
by the parameter $U$. The difference  $\gdir - \gind$
is in this approximation enhanced, due to a change of the character
of the lowest conduction band along the symmetry lines. Hybrid calculations
using HSE03 give a comparable \gdir\ and a modified dispersion
of both valence and conduction states close to the
Fermi energy, which reduces $\gdir - \gind$. The conduction
band minimum (CBM) within HSE03 is still located at $\Gamma$, but 
the difference between the CBM at L and $\Gamma$
gets significantly smaller. For sc-\GW, 
besides the further increase of the band gaps, the dispersion
of the bands is strongly affected by the many-body
effects. In fact, the \GW\ corrections exhibit an unusual
dispersion of around 1\,eV when looking at the different
${\bf k}$-points, displacing the CBM from
$\Gamma$ to L. We note that often in semiconductor physics one
assumes that the quasiparticle corrections can be modeled
by a rigid shift (the so-called scissor operator). From
our results it follows that one should refrain from using
this simple approximation for these important materials.
We can also conclude that hybrid calculations give a better
description of band dispersions than LDA+$U$, even if
the two approaches yield similar band gaps.

In conclusion, it is clear that the delafossite family exhibits
complex and unusual band gap physics that can
not be captured by standard theoretical approximations.
We found that the direct band gap is well reproduced
by the best many-body approaches if polaronic effects
are taken into account. We can expect that this situation, of a large gap that is reduced
substantially by polaronic effects, is quite general and is present in
many more materials that previously expected. 
In fact, the apparent good agreement between calculated gaps (with hybrid functionals or
\GWz) and experimental gaps for materials as simple and
widely studied as LiF can be accidental, as preliminary calculations confirm: the underestimation of the gap by 
these methods (the sc\GW\ gap is indeed 2\,eV larger than the experimental
and \GWz\ gap) is compensated by the neglect of large polaronic effects.
Furthermore, the modifications with respect to the LDA Kohn-Sham bands 
are strongly $\kv$-dependent,
which makes questionable the common practice of using a scissor operator.
The band dispersion obtained by hybrid functional calculations
is in between the LDA and sc-\GW dispersion, while the LDA+$U$ calculations
open up the gap but do not give a significant improvement of the band dispersion.
Finally, our calculations rule out the interpretation of the low energy features in
the absorption spectra as arising from a putative indirect
band gap. These structures should rather come from intrinsic
defects, as proposed in Refs.~\cite{pellicer06,robertson02}. However,
a complete understanding of the electronic
and excitation properties of these systems
will only be achieved, in our opinion, by a high-level theoretical
scheme (like sc-\GW) including defects and
effects from the lattice polarization in an ab initio framework.
Work along these lines is in progress.

We thank F. Bechstedt and A. Rubio for fruitful discussion.
Part of the calculations were performed at
the LCA of the University of Coimbra and at GENCI
(project x2009096017). SB acknowledges funding
from the European Community through the e-I3 ETSF project
(Contract \#211956), MALM from the Portuguese FCT (PTDC/FIS/73578/2006) and from the French
ANR (ANR-08-CEXC8-008-01), JV from an EDF/ANR CIFRE fellowship.

\ifx\mcitethebibliography\mciteundefinedmacro
\PackageError{apsrevM.bst}{mciteplus.sty has not been loaded}
{This bibstyle requires the use of the mciteplus package.}\fi


%
%
%
%
%


\begin{mcitethebibliography}{44}
\bibitem[{not()}]{note1}
\bibinfo{note}{The polaron constant is defined as $\alpha_{\textrm{P}} =
  (1/\epsilon_\infty - 1/\epsilon_0)(\hbar / 2 m a_{\textrm{B}}^2
  \omega_{\textrm{LO}})^{1/2} $, where $a_{\textrm{B}}$ is the Bohr radius,
  $\omega_{\textrm{LO}}$ the longitudinal optical frequency of the highest
  $E_u$ phonon mode, $\epsilon_\infty$ and $\epsilon_0$ the low frequency
  electronic and the static dielectric constants, respectively}\relax
\mciteBstWouldAddEndPuncttrue
\mciteSetBstMidEndSepPunct{\mcitedefaultmidpunct}
{\mcitedefaultendpunct}{\mcitedefaultseppunct}\relax
\EndOfBibitem
\end{mcitethebibliography}
\end{document}